\documentclass[aps,showpacs,twocolumn]{revtex4}
\usepackage{graphicx}
\def \bb{\mathbf}
\def \roty{\nabla_y \times}
\def \rotx{\nabla_x \times}

\def \e{\varepsilon}
\begin{document}
\title{Theory of mesoscopic magnetism in meta-materials}
\author{Didier Felbacq}
\affiliation{GES UMR-CNRS 5650\\
Universit\'{e} de Montpellier II\\
B‰t. 21, CC074, Place E. Bataillon\\
34095 Montpellier Cedex 05, France} 

\author{Guy Bouchitt\'e}
\affiliation{Laboratoire ANAM \\
Universit\'{e} de Toulon et du Var, BP 132 \\ 83957 La
Garde Cedex, France}
\begin{abstract}
We provide a rigorous theoretical basis for the artificial magnetic activity of meta-materials near resonances.
 Our approach is a renormalization-based scheme that authorises a completely general theory. The major result is
 an explicit expression of the effective permeability, in terms of resonant frequencies.
The theoretical results are checked numerically and we give applications of our theory 
to Left-Handed Media and to the solution of the Pokrovski-Efros paradox.
 
\end{abstract}
\maketitle

Photonic crystals are artificial devices, periodically structured, that exhibit photonic band gaps \cite{dowling}.
 Dielectric photonic crystals are considered in the optical domain, but metallic ones (or wire mesh photonic crystals)
  are also studied in the microwave or TeraHz range \cite{yablo,smithtera}.
It has been well established \cite{pendryplasmon} that, below a cut frequency, wire mesh photonic crystals 
behave as if they were homogeneous with a negative, frequency-dependent, permittivity
given by 
$\varepsilon_{\rm eff}=1-2\pi \gamma/
\left( \omega /c \right)^2$, where $\gamma=d^2 \log(d/r)$
(here $d$ is the period of the crystal and $r$ the radius of the wires) \cite{soukou}. 
For a  fequency below $\omega_p=\sqrt{2 \pi \gamma}\,c$, the homogenized
permittivity is negative and the propagation of waves is forbidden.
The homogenized permittivity represents the scattering behavior of the wire mesh photonic crystal
for large enough wavelengths, and explains why these structures present a photonic band
gap down to the null frequency (at least for infinitely conducting wires).
Recently, Pendry and co-authors suggested that it was possible to design photonic crystals with
non-magnetic materials that they possess an artificial magnetic activity \cite{pendry2}
and be described by an effective permeability. Basically, two geometries have been suggested: Split Ring Resonators
and dielectric fibers with a large permittivity \cite{pendry2,obrien}. It 
is believed that with these
geometries it is possible to obtain a negative permeability, 
and, by adding a wire mesh structure, to design a material with both a negative
permittivity and a negative permeability. Materials with these characteristics
do not seem to exist in nature, and therefore one tries to design them artificially 
(they are called "Left-Handed Materials"). 
They were studied theoretically long ago in a speculative and
quite fascinating work \cite{veselago} by Veselago. He showed that they
behave as if they had a negative index. Among other properties, Snell-Descartes law is reversed:
 at the interface between air and the material a beam is refracted on the same side of the normal. 
 These ideas have motivated a lot of works, 
both experimentally and numerically (in particular in \cite{smith}), and also
a lot of polemical issues \cite{val}.

 It seems however that, hitherto, there be no unified theoretical approach to this kind
of effective behavior. The method generally followed consists in characterizing
the scattering matrix of a basic resonator by means of the $\e$ and $\mu$ parameters, and
then deriving the effective parameters
without taking into account the coupling between each resonator \cite{muneg}. 
In the present work, we address
this problem by using a renormalization group analysis, which gives us a deep
insight into the phenomena and predicts that the existence of a negative $\mu$
is linked to internal resonances. In fact, the possibility of getting a negative permeability
is very different from the possibility of getting a negative permittivity: while
the negative $\e$ is obtained for low-frequencies (i.e. large wavelength
with respect to the wires constituting the crystal), the negative $\mu$ is obtained
in the resonant domain, and only for a rather small interval of frequencies. 
In particular, our approach explains the apparent paradox raised by Pokrovsky  al. \cite{pokro}, that, 
by embedding wires in a medium with negative $\mu$, one does not get a Left Handed
Medium. It gives also a complete analysis of the effective properties of wire mesh photonic
crystal, with a very high conductivity.
 Before going into the details of our study, we stress that we have tried
here to make a bridge between two domains that seem to be antagonist: that of the resonances and
that of homogenization.

\begin{figure}[h]
    \includegraphics*[width=6cm,height=4cm]{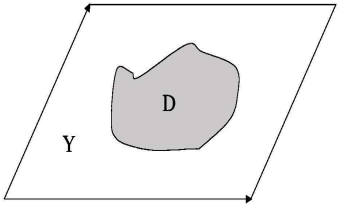}
    \caption{The basic cell of the photonic crystal}
    \label{schema}
 \end{figure}

In the following, we consider a $2D$ photonic crystal whose Wigner-Seitz cell $Y$ is given in 
fig. \ref{schema}. It is made of a dielectric rod (relative permittivity $\e_{i}$, cross-section $D$) embedded in a dielectric
matrix $\e_{e}$. When the contrast between $\e_{i}$ and $\e_{e}$ is substantial enough, there appear Mie resonances into
the highest index material. It was suggested some time ago \cite{moroz} that such internal
 resonances might result in the opening of forbidden gaps. 
Our point is to show that, near these resonances, the device behaves as if it had homogeneous 
electromagnetic parameters $\e_{h}$ and $\mu_{h}$.
 Of course, for this situation to be physically sounded, the resonant wavelengths should be larger
 than the period, otherwise the medium could not be described by homogeneous parameters. 
 That is why we request, as in \cite{pendry2}, that $\e_{i}$ be much higher then $\e_{e}$.
The method that we employ consists in changing ("renormalizing") the properties of the medium while
keeping the relevant physical phenomena, i.e. the resonances, unchanged. To do so, we choose
a small number $\tau <1$, and we proceed to the following operation, denoted 
${\cal R}$:
\begin{itemize}
\item We multiply the radius of the rods and the period by $\tau$, 
while maintening the domain $\Omega$ where the rods are contained constant 
(the number $N$ of rods is increased as $N \sim |\Omega|/\tau^{2}$).
\item We divide the permittivity $\varepsilon$ of the rods by $\tau^2$ 
(the optical diameter remains constant).
\end{itemize}
The wave is $p$-polarized so the induction field reads $\mathbf{B}(\bb{x})=u(\bb{x}) \bf{e}_{z}$, but the vectorial form will prove useful
for the analysis. We write $\mathbf{B}(\bb{x};{\cal R})$ 
and $\mathbf{E}(\bb{x};{\cal R})$ 
the fields scattered by the renormalized structure.
The point is to iterate this operation $n$ times and study the limit of 
$\mathbf{B}(\bb{x};{\cal R}^n)$ and $\mathbf{E}(\bb{x};{\cal R}^n)$ 
as $n$ tends to infinity. In order to do so,
we use a two-scale expansion of $(\bb{E,B})$:
\begin{equation}
\begin{array}{ll}
\bb{B}&=\bb{B}_0(\bb{x},\bb{x}/\tau^n)+\tau^n \bb{B}_1+...     \\
\bb{E}&=\bb{E}_{0}(\bb{x},\bb{x}/\tau^n)+\tau^n \bb{E}_1+...
\end{array}
\label{multiE}
\end{equation}
where the fields $\bb{E_{0},B_{0}}$ depend on both the real space variable $\bb{x}$ (the global variable)
and on the Wigner-Seitz cell variable $\bb{y}$ (the local variable). The fields are periodic with respect to $\bb{y}$.
Our point is to find the limit fields
$\bb{E_{0},B_{0}}$. The local variable is in fact a hidden one: it is an internal degree
of freedom. The true (observable) macroscopic fields $\left(\bb{E}_{h},\bb{B}_{h}\right)$ are the averages of 
the microscopic fields $\left(\bb{E}_{0},\bb{B}_{0}\right)$ over $Y$: 
\begin{equation}
\bb{B}_{h}(\bb{x})=\int_Y \bb{B}_0(\bb{x},\bb{y}) d\bb{y},
 \bb{E}_{h}(\bb{x})=\int_Y \bb{E}_0(\bb{x},\bb{y}) d\bb{y}.
\end{equation}
Although we do not find it relevant to present all the mathematical details, we believe that
it is important to offer the reader a general view of the method employed to get the limit fields.
A complete and rigorous mathematical derivation can be found in \cite{cras}

We analyze first the behavior of the fields with respect to the local variable.
That is, we wish to describe the microscopic behavior of the fields with respect to
their internal degrees of freedom. 
Using the expansion (\ref{multiE}) of the field, 
the $\nabla \times \cdot$ operator
is transformed into:
$$
\nabla \times \cdot \longrightarrow \nabla_x \times \cdot+
\tau^{-n} \nabla_y \times \cdot
$$
(we have to make explicit on what variables the derivations operate because there are two
sets of variables).
Plugging these expressions into Maxwell system and identifying the terms that corresponds to the same power of $\tau^n$ 
we obtain the following system for the microscopic electric field:
\begin{eqnarray}
\nabla_y \times \bb{E}_0&=0  \hbox{ on } Y \, , \,
\nabla_y \cdot \bb{E}_0&=0  \hbox { on } Y \setminus D 
\label{electro}
\end{eqnarray}
Besides: $\bb{E}_0=0 \, \hbox{ on } D $ and $\bb{E}_1=0  \, \hbox{ on } Y\setminus D$.
This system is of electrostatic type: $\bb{E}_0$ does not depend on the microscopic induction field
nor does it depend upon the wavelength. As a matter of fact, on $Y\setminus D, \,\, \bb{E}_{0}$ does not depend 
upon the variable $\bb{y}$, as it can be deduced from system (\ref{electro}).
Let us now turn to the magnetic field. The system satisfied by $\bb{B}_0$ is of an entirely different nature:
\begin{equation}
\begin{array}{rll}
\roty \bb{B}_0 &=-i \omega \e_i \bb{E}_1 & \hbox{ on } Y \\
\roty \bb{E}_1&=i \omega \bb{B}_0 & \hbox{ on } D 
\end{array}
\label{sysb}
\end{equation} 
We have obtained a microscopic Maxwell system that describes the 
microscopic behavior of the fields. It can be seen
that $\bb{E}_1$ gives indeed a first order expansion of the field
inside $D$: it replaces $\bb{E}_0$ which is null there. 
Let us now use the fact that the fields are polarized. We write:
$\bb{B}_{0}(\bb{x})=u_{0}(\bb{x}) \bb{e}_{z}$. Plugging this expression into
 system (\ref{sysb}) shows that the
magnetic field is independent of $\bb{y}$ on $Y\setminus D$. Next, by combining the 
equation is system (\ref{sysb}), we find that:
\begin{equation}
\Delta_{y} u_0 +k^2 \e_i u_0=0  \hbox{ on } D \,, \,
u_0=\hbox{cst}   \hbox{ on } Y\setminus D
%
\end{equation}
We deduce from this system that the microscopic induction field is linked to the macroscopic one by:
$u_0(\bb{x},\bb{y})=(m(\bb{y})/ \mu_{h}) \, u_h(\bb{x})$
%
where $m$ satisfies:
\begin{equation}
\Delta_{y} m+k^2\e_{i} m=0  \hbox{ on } D \,,\,
m=1  \hbox{ on } Y\setminus D
%
\label{mu}
\end{equation}
and $\mu_{h}$, which shall be interpreted below as a relative permeability,
 is the mean value of $m :\,\mu_{h}=\int_{Y} m(\bb{y}) d\bb{y}$. 
Up to now, we have clarified what happens at the microscopic scale.
The point is now to derive the equations that are satisfied by the macroscopic fields.
  The propagation equations read, for $\bb{y} \in Y\setminus D$:
\begin{equation}
\begin{array}{ll}
\rotx \bb{B}_0+\roty \bb{B}_1&=-i \omega \e_{0}\e_e \bb{E}_0 \\
 \rotx \bb{E}_0+\roty \bb{E}_1&=i \omega \bb{B}_0
 \end{array}
 \label{microH}
\end{equation}
In the first line, we recognize the Maxwell-Amp\`ere equation with the extra-term
$\roty \bb{B}_{1}$. This term is homogeneous to an electric displacement field,
and it represents the polarisation due to the presence of the scatterers. Indeed, 
in the long wavelength regime, the emission diagram of a fiber is that
of a dipole (for the $p$-polarization). As a consequence, the whole set of fibers
that constitutes the photonic crystal behaves as a set of coupled dipoles,
producing a possibly anisotropic permittivity tensor. 
More precisely, as $u_{0}$ does not depend on $\bb{y}$
on $Y\setminus D$, we obtain the following system satisfied by $u_{1}$:
\begin{equation}
\Delta_{y} u_{1}=0  \hbox{ on } Y \setminus D \, , \,
\frac{\partial u_{1}}{\partial n}=-{\mathbf n}\cdot \nabla_{y} u_{0}
\hbox{ on } D,
\end{equation}
where $\bb{n}=(n_{1},n_{2})$ is the normal to $D$.
This system implies a linear relation between $u_{1}$ and
$u_{0}$ of the form:$\, \nabla_{y} u_{1}={\cal P}(\bb{y}) \nabla_{y} u_{0}$
where:
\begin{equation}
 { \cal P}(\bb{y})=\left (\begin{array}{lr}
1+\frac{\partial w_{1}}{\partial x_{1}} & \frac{\partial w_{2}}{\partial x_{1}} \\
\frac{\partial w_{1}}{\partial x_{2}}  & 1+\frac{\partial w_{2}}{\partial x_{2}}  
\end{array}\right)
\end{equation}
and $w_{i}$ satisfies:
\begin{eqnarray}
\Delta w_{i} =0  \hbox{ on }  Y\setminus D  \, ,\,
\frac{\partial w_{i}}{\partial n} =-n_{i}  \hbox{ on }  \partial D 
\label{annex}
\end{eqnarray}
It is not difficult to see \cite{koz,wave} that $A_{h}=\int_{Y} {\cal P}(y)dy$
is the inverse of the effective permittivity tensor 
$\e_{h} (\,=A^{-1}_{h})$ of
the photonic crystal. The effective macroscopic equation can now be obtained
by averaging system (\ref{microH}) on $Y \setminus D$:
 \begin{equation}
 \nabla \cdot (\e^{-1}_{h}  \nabla (\mu_{h}^{-1}  u_{h}))+k^2 u_{h}=0
 \end{equation}
The macroscopic behavior of the system is characterized by an effective
permittivity tensor $\e_{h}$ and an effective permeability $\mu_{h}$ . 
This shows that the system exhibits an artificial magnetic activity.
There are two huge differences between the effective permittivity and the effective permeability:
the permittivity can be a matrix, hence the medium can be anisotropic, whereas the effective
permeability is always a scalar, therefore no anisotropic permeability can be obtained. Second,
the permittivity is not frequency dependent, it is a static permittivity, whereas the permeability
depends on the frequency.
Let us give a closer look at the system of equations that defines 
the effective permeability $\mu_{h}$. As it stands in (\ref{mu})
it is just a partial differential equation problem. Under this form, its physical meaning does
not appear clearly. To make it more explicit, let us recast it into an eigenvalue problem. This will help
us understanding the underlying physics of what might look, at this stage, as
a rather abstract result. The system (\ref{mu}) has a unique solution only if there is no function $\psi$
such that $\psi$ is null on $Y\setminus D$ and $\psi$ satisfies the same Helmholtz equation
on $D$. Otherwise $m+\psi$ would still be a solution of (\ref{mu}). Following spectral theory
\cite{kato}, we denote $H=-\e_{i}^{-1}\Delta $ and we look for functions $\Phi$ satisfying
the eigenvalue problem:
\begin{equation}
\Phi=0 \hbox{ on } Y \setminus D \, , \,
H\Phi=k^2 \Phi  \hbox{ on } D.
\label{eigen}
\end{equation}
We get a set of eigenvalues $k_n^2$
and a set of eigenfunctions $\left|\Phi_n \right >$. The physical meaning
of these eigenvalues can be understood by going back to the unrenormalized
initial fiber, with permittivity $\e_{i}$. This fiber alone 
possesses resonant frequencies. They correspond to modes that
are strongly localized inside the fiber. However, when there is 
a large number of fibers these resonances are slightly shifted due
to coupling, and these resonances are furthermore modified by the
renormalization process. That is precisely what the eigenvalues of
problem (\ref{eigen}) are: the renormalized Mie frequencies of the fibre.

For a given wavevector $k^2$,  we look for a solution $m$
by expanding it on the
basis $\left|\Phi_n \right >$, by noting that $m-1$ is null on $Y\setminus D$:
$\,
m(\bb{y})=1+\sum_n m_n \left|\Phi_n \right >.
$
The coefficients are obtained by inserting this expansion in (\ref{mu}).
 We get, after averaging, the effective permeability $\mu_{h}=\left<1|m\right>$ 
under the form:
\begin{equation}
\mu_{h}(k)=1+ k^2 \sum_{n} \frac{\left|\left<\Phi_n | 1 \right >\right|^2}{k_{n}^2-k^2}
\end{equation}
We have obtained a completely general expression for the effective permeability. It relies
on the cavity modes of the fibre only.
In the vicinity of a resonance $k_{n}^{2}$, we have: 
$\mu_{h} \sim 1-k_{n}^{2}\left|\left<\Phi_n | 1 \right >\right|^2 (k^2-k_n^2)^{-1}$
which shows, in complete generality, that the permeability exhibits
anomalous dispersion near the resonances, and becomes negative there. 
It should also be noted that only the eigenfunctions with non-zero mean value contribute. This
is due to the fact that we have only kept the first order terms in the expansions (\ref{multiE}).

Let us give an explicit computation in case of a square fiber. The
 derivation is rather straightforward, and follows closely that of the well-known $TE$ modes
in waveguides with square section.
The eigenfunctions are $\Phi_{nm}({\bf y})=2 \sin(n\pi y_{1})\sin(m\pi y_{2})$ and the 
corresponding eigenvalues are: $k_{nm}^{2}=\pi^{2}(n^2+m^2)$ 
The expansion of $m$ on this basis leads to the following effective permeability:
\begin{eqnarray}
\mu_{h}(k)=1+\frac{64 a^2}{\pi^{4}}\sum_{(n,m)\small{ odd }}
\frac{k^{2} }{n^{2}m^{2}(\tilde{k}_{nm}^2-k^2)}
\label{muef}
\end{eqnarray}
 where $\tilde{k}_{nm}^2=k_{nm}^{2}/a^2\e_{i}$.
Let us now turn to some numerical applications. First, we note that our analysis
is supposed to work when there are Mie resonances at wavelengths large with respect 
to the period of the crystal. This was the situation described in
\cite{obrien}, where $\varepsilon=200+5i$. We choose this
value for our own numerical computations, the point being to test the
validity of the theory.
 \begin{figure}[h]
    \includegraphics*[width=7cm,height=6cm]{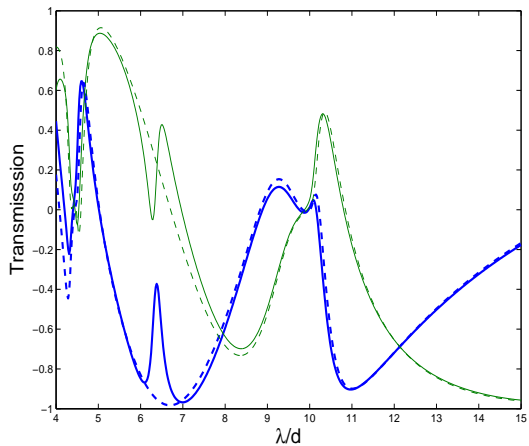}
    \caption{Real (bold lines) and imaginary (thin lines) parts of the transmission for the meta-material (solid lines) and the homogenized
    material (dashed line).}
    \label{trans}
 \end{figure}
Using a rigorous diffraction code for gratings \cite{neviere}, we plot the transmission spectrum 
(dashed line in fig. \ref{trans}) for a stack of $3$ diffraction gratings made of square rods.
 There is a band gap for $\lambda/d$ between $8$ and $12$, 
due to a Mie resonance. In order to test our results, we plot the transmission
spectrum of a homogeneous slab (solid line fig. \ref{trans}) with parameters $\e_{h}=1.7$ (this value is obtained
numerically from the resolution of problem (\ref{annex})) and $\mu_{h}$ 
given in (\ref{muef}).
We see in fig. \ref{trans} that both curves fit very well, indicating that, although the
wavelength is not that large, the whole photonic crystal behaves as a homogeneous magnetic
material. The discrepancy that is seen around $\lambda/d=6.5$ is due to the presence
of a Mie resonance of null mean value that is not taken into account in our theory. Only
by expanding the fields to the second order could we incorporate this resonance in our 
global result.

Finally, let us use these results to analyze recent problem.
In \cite{pokro}, Pokrosky and al. showed that
it was not possible to design a negative index medium by embedding
metallic wires into a matrix with a negative $\mu$, whereas the 
converse is possible. This can be explained in the following manner:
the negative permittivity is obtained as a macroscopic effect, by
which we mean that it is an interference effect and not an effect
that takes place at the scale of the microscopic cell only. In much a different
way, the negative permeability is obtained as a purely local effect,
that happens at the scale of the microscopic cell. Therefore, for this
effect to occur, no strong coupling between the fibers is requested, 
the coupling has just to be sufficient enough that the incident field can reach
the fibers by tunnel effect. In our model, the propagation equation of the structure 
is obtained immediately by replacing $\e_{e}$ by $-\e_{e}$.
Then, near the regions of anomalous dispersion, both parameters
are negative and the propagation equation is the usual Helmholtz
equation. Consequently, the field can propagate. On the contrary,
for metallic wires in a medium with negative $\mu$ the propagation equation
reads \cite{wave,optlett}: $ \Delta u+k^{2}\mu (1-2\pi \gamma/\mu k^{2})u=0$, which
leads to evanescent waves.


We have given in this work a theory of the mesocopic magnetism in metamaterials. We have 
shown that it was possible to give a homogenized
description of a heterogeneous device in the resonance domain. To do so,
we have used a renormalization approach that shows that two scales
should be distinguished: a microscopic one and a macroscopic one. 
We have shown that the artificial, mesoscopic, magnetism is due to microscopic magnetic moments
induced by geometric resonances. So far, the analysis works for
high permittivities, but we stress that there are inner resonances
in gratings for much lower contrasts as well \cite{mc}. Therefore, we do believe
that the same physics can be found in the optical range of wavelengths.


\begin{thebibliography}{}
\bibitem{dowling} http://phys.lsu.edu/~jdowling/pbgbib.html
\bibitem{yablo} D. F. Sievenpiper, M. E. Sickmiller, and E. Yablonovitch, Phys. Rev. Lett. {\bf 76}, 2480 (1996).
\bibitem{smithtera} T. J. Yen et al., Science {\bf 303}, 1494 (2004).
\bibitem{pendryplasmon} J. B. Pendry, A. J. Holden, W. J. Stewart and I. Youngs, Phys. Rev. Lett. {\bf 76}, 4773 (1996).
\bibitem{soukou}There are some variants with different constants, see 
P. Marko, C. M. Soukoulis, Opt. Lett. {\bf 28},  846  (2003).
\bibitem{pendry2} Pendry J B, Holden A J, Robins D J and Stewart W J, 
IEEE Trans. Microw. Theory Tech. {\bf 47}, 2075 (1999).
\bibitem{obrien} S. O'Brien and J. B. Pendry, J. Phys.: Condens. Matter{\bf14}, 14035 (2002).
\bibitem{veselago} V. G. Veselago, Sov. Phys. Usp. {\bf 10}, 509 (1968). 
\bibitem{smith}R. A. Shelby, D. R. Smith, S. Schultz, Science {\bf 292}, 77 (2001).
\bibitem{val}P. M. Valanju, R. M. Walser, and A. P. Valanju, Phys. Rev. Lett. {\bf 88}, 187401 (2002).
\bibitem{muneg} M. Shamonin, E. Shamonina, V. Kalinin, and L. Solymar, J. Appl. Phys. 95, 3778 (2004). Ê
\bibitem{pokro} A. L. Pokrovsky and A. L. Efros, Phys. Rev. Lett. {\bf 89}, 093901 (2002).
\bibitem{moroz} A. Moroz, A. Tip, J. Phys.: Condens. Matter {\bf 11}, 2503 (1999).
\bibitem{kato} T. Kato, Perturbation theory for linear operators, Springer-Verlag, Berlin, 1995.
\bibitem{cras} G. Bouchitt\'e, D. Felbacq, C. R. Acad. Sc. Paris, Ser. I {\bf 339}, 377 (2004).
\bibitem{koz} V.V. Jikov, S.M. Kozlov, and O.A. Oleinik,
 Homogenization of Differential Operators and Integral Functionals, Springer-Verlag, NY, 1994. 
\bibitem{wave} D. Felbacq, G. Bouchitt\'e, Waves in Random Media  {\bf 7}, 245 (1997).
\bibitem{neviere} M. Nevi\`ere, E. Popov, Light Propagation in Periodic Media: Differential Theory and Design,
Marcel Dekker, 2002.
\bibitem{optlett} D. Felbacq, G. Bouchitt\'e, to be published in Optics Letters.
\bibitem{mc} M. C. Larciprete, D. Felbacq, in preparation.
\end{thebibliography}
\end{document}